# Space Weather Prediction from the Ground: Case of CHAIN

**Daikichi SEKI[1,2], Satoru UeNo[1], Hiroaki Isobe[3], Kenichi Otsuji[1], Denis P. Cabezas[1], Kiyoshi Ichimoto[1], Kazunari Shibata[1], and CHAIN team[1].**

1. **Kwasan and Hida Observatories, Kyoto University, Japan**
2. **GSAIS*, Kyoto University, Japan**
3. **Faculty of Fine Arts, Kyoto City University of Arts, Japan;**

*Graduate School of Advanced Integrated Studies in Human Survivability

E-mail (seki@kwasan.kyoto-u.ac.jp)

**Abstract**

In this article, we insist on the importance and the challenges of the prediction of solar eruptive phenomena including flares, coronal mass ejections (CME), and filament eruptions fully based on the ground-based telescopes. It is true that satellites' data are indispensable for the space weather prediction, but they are vulnerable to the space weather effects. Therefore, the ground-based telescopes can be complementary to them from the viewpoint of space weather prediction.

From this view point, one possible new flare prediction method that makes use of H-alpha, red wings, and blue wings images obtained by the SDDI/SMART, the ground-based telescope at Hida Observatory, is presented. And in order to show the possibility for the actual operation based on that method, the recent progress of CHAIN project, the international observation network, is mentioned in terms of their outcomes and capacity buildings.

**Keywords:** space weather prediction, ground-based telescopes, CHAIN project, multi-wavelength H-alpha imaging, filament eruption

# 1. INTRODUCTION

Space weather is the disturbances in the interplanetary plasmas and in the interplanetary magnetic fields mainly caused by the solar activity. It is known that space weather has potential risks to damage human technologies in forms of satellite anomalies, satellite air drag, single events, radiation exposure, and huge blackout (Schrijver et al., 2015). As a global navigation satellite system like GPS is now one of the essential social infrastructures, human beings will be getting more and more dependent on space technology in future. Therefore, in order to mitigate and prevent space weather disaster, it is highly significant to predict when and how large space weather events will happen.

Currently, space weather prediction is mainly done by using space-borne instruments such as the Atmospheric Imaging Assembly (Lemen et al., 2012) and the Helioseismic and Magnetic Imager (Scherrer et al., 2012) on the Solar Dynamics Observatory, the Large Angle Spectroscopic Coronagraph (Brueckner et al., 1995) on the Solar and Heliospheric Observatory, and Geostationary Operational Environmental Satellite series. There are basically 2 reasons for this situation. One is that they can see the sun in soft and hard X-ray and extreme ultraviolet which we can never observe from the ground due to the air and are essential to see the magnetic fields, which give us the keys to predict solar flares. The other is that they can continue observation regardless of time and terrestrial weather. However, it does not mean that the ground-based telescopes are no longer useful in terms of space weather prediction. Firstly, compared to the satellites, the ground-based telescopes are cheap, and developing countries can also introduce the instruments. Secondly, the ground-based telescopes do not have any effects of space weather, while artificial satellites could be damaged by severe space weather events. Therefore, the ground-based telescopes can be used as backup, and even if a satellite observing in H-alpha and its wings should launch, they would still have some complementary roles in case of satellites' failure by a huge solar flare.

In this article, we present the possibility of space weather prediction, especially of predicting solar explosive phenomena, by the ground-based telescopes, with reference to the recent progress of the Continuous H-Alpha Imaging Network (CHAIN) project (UeNo et al. 2007) worked by Kyoto University. In the next section, a possible candidate of prediction systems will be introduced. In the first subsection, we will give a method of prediction with reference to the recent suggestion in Seki et al. (2017), and in the second subsection, from a hardware point of view the recent progress of CHAIN will be reported. And the 3 challenges we are facing will be mentioned in the last section.

# 2. PREDICTION BY H-ALPHA IMAGE

There are several observational methods from the ground such as a coronagraph, a magnetogram, a continuum light observation, and a H-alpha observation. At Hida Observatory in Kyoto University, there is a powerful instrument observing the sun in H-alpha line and its wings called Solar Dynamics Doppler Imager (SDDI) installed on Solar Magnetic Activity Research Telescope (SMART) (Ichimoto et al., 2017).

## 2-1  Prediction of a Filament Eruption by H-alpha Images (Seki et al. (2017))

SDDI has been conducting a routine observation since 2016 May 1. It takes the solar full-disk images of 73 channels at every 0.25 Å from the H-alpha line center −9.0 Å to the H-alpha line center +9.0 Å, i.e., at 36 positions in the blue wing, the H-alpha line center, and 36 positions in the H-alpha red wing. Each image is obtained with a time cadence of 15 seconds and a pixel size of about 1.2 arcsec (Ichimoto et al., 2017). When the weather permits, it continuously monitors the Sun during the daytime in Japan for about 10 hr. Details on the instrument and examples of images and line profiles can be found in Ichimoto et al. (2017).

Making the most of the small gap between channels of 0.25 Å and the good time cadence of 15 sec, Seki et al. (2017) deduced automatically the unprecedented detailed line-of-sight velocities of the filament which erupted around 4:00UT on 2016 November 5th by utilizing Beckers' cloud model (Beckers, 1964; Morimoto and Kurokawa, 2003a,b)) (See Figure 1). The brief explanation of the method is as follows;

1. determining the positions of the pixels whose intensities are lower than the subtraction between the average intensity of the full sun and the double of its standard deviation for each wavelength image,
2. getting all the positions together to obtain a binary image whose pixels at the same positions have 1 and the other pixels have 0,
3. smoothing the binary image by taking the average of 5" x 5" pixels around each pixel,
4. conducting "erosion-dilation-dilation-erosion" process in the binary image ("erosion" and "dilation" are fundamental techniques in the field of morphological image processing) to get the "mask" which cover the most of the pixels inside a filament, and
5. calculating the line-of-sight velocities of a filament inside the "mask" based on Beckers' cloud model.

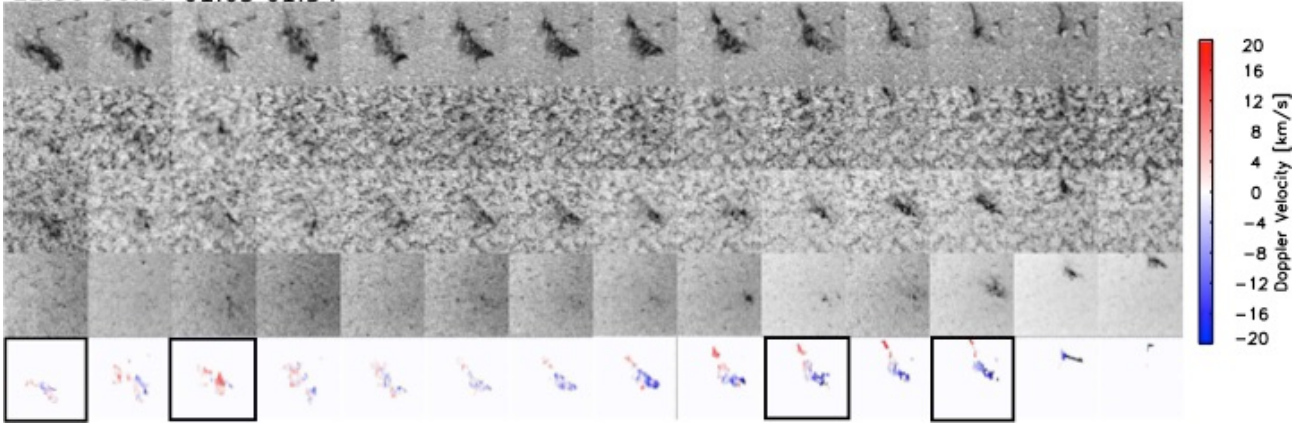

**Figure 1.** Time series of H-alpha images at the line center, at +0.5 Å, at −0.5 Å, and at −1.0 Å (the top four rows), and of the images of LOS velocity of the filament (bottom row). (Seki et al., 2017)

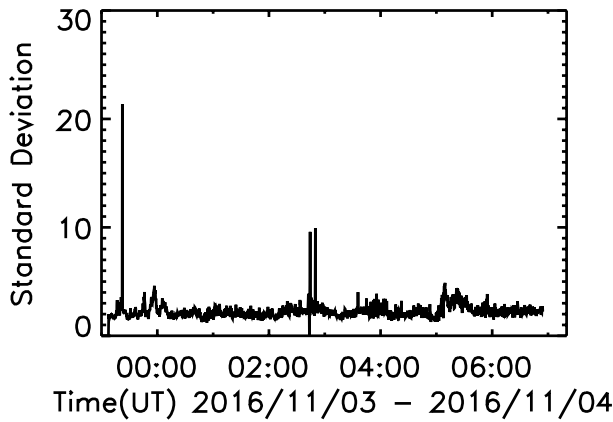
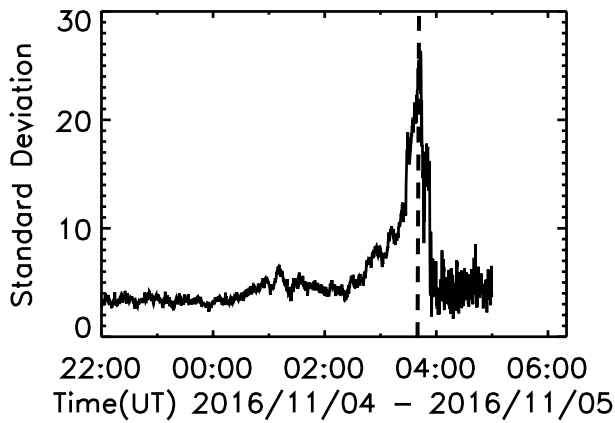

As a result, by tracking the standard deviation of line-of-sight velocities for each velocity map, we found that it increased sharply around 1 hour before eruption. Figure 2 shows the time transition of the standard deviation. On the previous day of the eruption, the standard deviation was almost constant around 2-3 km s$^{-1}$. However, on the next day (the day of the eruption) it slightly increases to 3-4 km s$^{-1}$ and stays constant until around 0:30UT. Then it gradually increases at a rate of 1.1 m s$^{-2}$ until it peaks around 1:10UT, and it starts to increase again sharply at a rate of 2.8 m s$^{-2}$ around 2:30UT, which corresponds to around 1 hour before eruption. In this case, accompanied by a filament eruption, B-class flare started around 4:00UT and the X-ray flux peaked around 4:40UT, and also weak CME occurred. Therefore, this work suggests that if we track the standard deviation of the line-of-sight velocities of a filament, we may be able to predict solar explosive phenomena around 1 hour before it happens only by using ground-based telescopes' data.

From the operational point of view, it should be noted that in 2017 September we started the automatic detection and calculation of line-of-sight velocity of a filament during the observation every day. The deduced data has been stored at Hida Observatory and everyone can access it via its web site http://www.hida.kyoto-u.ac.jp/SMART/SDDI/cloudmodel/. Combining this system and the result of Seki et al. (2017), it must be easy to monitor filaments during observation and to create an automatic alerting system of filament eruptions.

## 2-2  CHAIN project
### 2-2-1  Introduction of CHAIN

We have been working Continuous H-Alpha Imaging Network (CHAIN) project, which is to create a world-wide observational network with ground-based solar telescopes observing in multiple wavelengths including H-alpha line center, red wing and blue wing for the purpose of 24-hour continuous observation of the three-dimensional velocity fields of filament eruptions and shock-wave structures on the whole solar surface (See No.6 of List of ISWI Projects at http://www.iswi-secretariat.org)(UeNo et al., 2014). Actually, there are other H-alpha network such as the Global Oscillation Network Group (GONG) (Harvey et al., 2011) and the Global H-alpha Network (GHN) (Steinegger et al., 2000). However, it is only this network group all over the world that is observing not only H-alpha center but also its red and blue wings.

There are 3 purposes for this project;
1. reinforcement of observations of the solar activity by formation of an international network of ground-based solar observations for 24 hours continuously,
2. observation and study of filament eruptions, shock waves (Moreton waves) with solar flares and variation of UV radiation on the full-disk of the Sun in order to understand and predict the change of space-weather environment from the Sun to the Earth,
3. international spread, academic exchange and promotion of the space-weather research including developing countries.

From the space weather point of view, there are 2 important aspects of CHAIN project, (1) constructing ground-based 24-hour space weather prediction system by creating an international solar observational network with ground-based telescopes and (2) capacity building including technical and scientific training and space weather education through installing and operating the ground-based solar telescopes.

### 2-2-2  Outcomes of CHAIN

As its outcomes so far, 3 countries including Japan, Peru, and Saudi Arabia are cooperating, and 17 papers related to CHAIN have been published (See Table 1). The contents of 7 papers are related to CHAIN project itself. The brief summaries of the contents are as follow:

- UeNo et al. (2007): It is the first paper related to CHAIN, and they explained about the project and their plan.
- Ishitsuka et al. (2007): They introduced their plan to refurbish solar observing stations in Peru and mentioned about installing FMT from Kyoto University in that context.
- UeNo et al. (2009): This paper published in Data Science Journal is explaining about data archive and observing system of CHAIN, and they claimed the necessity of improving the information technology.
- UeNo et al. (2010): With the introduction of CHAIN project, they mentioned that Algeria is one of the best candidates of this project and explained their plan of investigation.
- Seghouani (2010): He claimed the necessity of a new astronomical observatory in Algeria and mentioned about Dr. UeNo's visit and investigation.
- Ishitsuka et al. (2014): They showed the summaries of capacity building, observed data, and scientific results brought by the FMT installed in Peru as part of CHAIN project.
- UeNo et al. (2014): They showed the progresses about international collaboration and academic exchange of CHAIN from 2010 to 2013.

In the other 10 papers, CHAIN data are used for their scientific researches (See Table 1). Brief explanations are shown below,

- Nagashima et al. (2007): They used SMART H-alpha data to investigate the position and the active motion of 2 filaments.
- Narukage et al. (2008): SMART data were used to discover 3 Moreton waves on 2005 August 3.
- Asai et al. (2009): SMART data were used to show the temporal evolution of the AR NOAA 10798 in H-alpha.
- Zhang et al. (2011): FMT multi-wavelengths data were used to study the statistical properties of propagating Moreton waves.
- Asai et al. (2012): SMART data were used to detect a Moreton wave on 2011 August 9 and associated filament/prominence oscillations.
- Ishii et al. (2013): A new high-speed imaging system for solar flares installed on SMART was introduced.
- Shen et al. (2014): SMART data were used to calculate the Doppler velocity of a filament to investigate its oscillation property.
- Ichimoto et al. (2017): A new $H\alpha$ and its wings imaging instrument, the Solar Dynamics Doppler Imager, was introduced.
- Cabezas et al. (2017): FMT data obtained at the National University San Luis Gonzaga of Ica, Peru, were used to derive the 3-dimensional velocity field of a filament eruption associated with a M-class flare on 2011 February 16.

- Seki et al. (2017): SMART/SDDI data were used for analyzing the amplitude of the small-scale motion of the filament which erupted on 2016 November 5.

As for the capacity building, 7 lectures, 4 scientific educations, 2 technical trainings, and 5 data-analysis workshops have been held (See Table 2) for these 10 years.

Table 1. Published Papers related to CHAIN project

| Year | Title | Authors | Journal |
|---|---|---|---|
| 2007 | CHAIN-Project and Installation of the Flare Monitoring Telescopes in Developing Countries | UeNo, S., Shibata, K., Kimura, G. at al. | Bulletin of the Astronomical Society of India 35, 697. |
| 2007 | Triggering Mechanism for the Filament Eruption on 2005 September 13 in NOAA Active Region 10808 | Nagashima, K., Isobe, H., Yokoyama, T. et al. | Astrophysical Journal 668, 533. |
| 2007 | A solar observing station for education and research in Peru | Ishitsuka, J., Ishitsuka, M., Avilés H. T. et al. | Bulletin of the Astronomical Society of India 35, 709 |
| 2008 | Three Successive and Interacting Shock Waves Generated by a Solar Flare | Narukage, N., Ishii, T. T., Nagata, S. et al. | Astrophysical Journal Letters 684, L45 |
| 2009 | The CHAIN- Project and Installation of Flare Monitoring Telescopes in Developing Countries | UeNo, S., Shibata, K., Kitai, R. et al. | Data Science Journal 8, 30 |
| 2009 | Evolution of Anemone AR NOAA 10798 and the Related Geo- Effective Flares and CMEs | Asai, A., Shibata,K., Ishii, T. T. et al. | Journal of Geophysical Research 114, A00A21 |
| 2010 | Un Observatoire dans la Région de Aurès | Seghouani, N. | African Skies 14, 44 |
| 2010 | Continuous H- alpha Imaging Network Project (CHAIN) with Ground-based Solar Telescopes for Space Weather Research | UeNo, S., Shibata, K., Ichimoto, K. et al. | African Skies 14, 17 |
| 2011 | Propagation of Moreton Waves | Zhang, Y., Kitai, R., Narukage, N. et al. | Publications of the Astronomical Society of Japan 63, 685 |

| Year | Title | Authors | Publication |
|------|-------|---------|-------------|
| 2012 | First Simultaneous Observation of an Hα Moreton Wave, EUV Wave, and Filament/Prominence Oscillations | Asai, A., Ishii, T. T., Isobe, H. et al. | Astrophysical Journal Letters 745, L18 |
| 2013 | High-Speed Imaging System for Solar-Flare Research at Hida Observatory | Ishii, T. T., Kawate, T., Nakatani, Y. et al. | Publications of the Astronomical Society of Japan 65, 39 |
| 2014 | International Collaboration and Academic Exchange of the CHAIN Project in this Three Years (ISWI Period) | UeNo, S., Shibata, K., Morita, S. et al. | Sun and Geosphere 9, 97 |
| 2014 | Within the International Collaboration CHAIN: a Summary of Events Observed with Flare Monitoring Telescope (FMT) in Peru | Ishitsuka, J., Asai, A., Morita, S. et al. | Sun and Geosphere 9, 85 |
| 2014 | A Chain of Winking (Oscillating) Filaments Triggered by an Invisible Extreme-ultraviolet Wave | Shen, Y., Ichimoto, K., Ishii, T. T. et al. | Astrophysical Journal 786, 151 |
| 2017 | A New Solar Imaging System for Observing High-Speed Eruptions: *Solar Dynamics Doppler Imager* (SDDI) | Ichimoto, K., Ishii, T. T., Otsuji, K. et al. | Solar Physics 292, 63 |
| 2017 | Increase in the Amplitude of Line-of-sight Velocities of the Small-scale Motions in a Solar Filament before Eruption | Seki, D., Otsuji, K., Isobe, H. et al. | Astrophysical Journal Letters 843, L24 |
| 2017 | "Dandelion" Filament Eruption and Coronal Waves Associated with a Solar Flare on 2011 February 16 | Cabezas, D. P., Martínez, L. M., Buleje Y. J. et al. | Astrophysical Journal 836, 33 |

Table 2. Capacity Building Activities

| Type | Date | Place | Country |
|---|---|---|---|
| Lecture | January 2007 | Ica (National Ica Univ.) | Peru |
| | January 2007 | Lima(Instituto Geofísico del Perú) | Peru |
| | May 2008 | Bouzaréah (Centre de Recherche en Astronomie, Astrophysique et Geophysique) | Algeria |
| | June 2008 | Ica | Peru |
| | March 2010 | Ica | Peru |
| | May 2011 | Riyadh | Saudi Arabia |
| | August 2015 | Riyadh | Saudi Arabia |
| Scientific Education | June 2010 | Ica (National Ica Univ.) | Peru |
| | October 2010 | Ica (National Ica Univ.) | Peru |
| | November 2010 | Ica (National Ica Univ.) | Peru |
| | October 2015 | (King Saud Univ.) | Saudi Arabia |
| Technical Training | January 2007 | Ica (National Ica Univ.) | Peru |
| | July 2009 | Hida (Hida Observatory) | Japan |
| Data-analysis workshop | November 2010 | Ica | Peru |
| | July 2011 | Hida (Hida Observatory) & Mitaka (National Astronomical Observatory of Japan) | Japan |
| | March 2013 | Hida (Hida Observatory) | Japan |
| | March 2015 | Kyoto (Kwasan Observatory) | Japan |
| | February 2017 | Kyoto (Kyoto Univ. & Kwasan Observatory) | Japan |

## 3. CHALLENGES FOR THIS TRIAL
### *3-1 Terrestrial Weather & Climate*
One of the weakest points of the ground-based telescopes is the influence of terrestrial weather. As for CHAIN in the current situation, if one of the institutions cannot observe the sun, there is no other observatory that can compliment the data. Moreover, FMT in Saudi Arabia has another effect from its nature. Because of its climate, the high temperature, it reaches around 50 degrees Celsius in the observation dome in the noon on summer. It is not expected for the filter installed on the telescope to become over 45 degrees Celsius so that it makes errors in observing wavelengths.

### 3-2 Smooth Data Access

Actually, all the 3 stations do observe the sun as long as it can, but as for the data sharing, some difficulties still exist. Currently internet line speed is very slow and it takes more time to send each station's observing data to Japan than for them to observe the sun. Because of this situation, we mainly gather the data at Japan not by internet but by currying HDD.

### 3-3 Applicability of the Suggested Prediction Method

The last challenge is the applicability of the new prediction method presented by Seki et al. (2017). In the paper, we presented one possibility to predict solar explosive phenomena, but the feature was investigated for only one event. Therefore, as the future work, it should be confirmed statistically whether the increase in the amplitude of line-of-sight velocity of small-scale motions in a solar filament can be seen in almost all the filaments before their eruption. In addition, if we actually try to use CHAIN as the data source of the prediction, we also should confirm whether FMT data can be used for this prediction method, because the less amount of observing wavelengths leads to the less accurate estimation of line-of-sight velocities. (SDDI has been observing at every 0.25 Å from H$\alpha$ line - 9 Å to H-alpha line + 9 Å every 15 seconds, while FMTs have been observing at 3 or 5 wavelengths, H-alpha line center and H-alpha line ± 0.8 Å in Peru, or H-alpha line center, H-alpha line ± 0.6 Å, and H-alpha line ± 1.2 Å in Saudi Arabia, in every 20 seconds.) Therefore, it should be checked whether the suggested precursor can also be detected from FMT data or not.

## Acknowledgement

We appreciate the daily routine observation and the maintenance and development of the instruments by the staff of Hida observory members. We are also grateful to the anonymous referees for their fruitful comments that greatly improved our work.